\documentclass[prb,preprint,showpacs,preprintnumbers,amsmath,amssymb]{revtex4}
\usepackage{graphicx}
\begin{document}
\title{Gap opening in topological-defect lattices in graphene}
\author{Joice da Silva-Ara\'ujo}
\author{H. Chacham}
\author{R. W. Nunes}\thanks{corresponding author}
\email{rwnunes@fisica.ufmg.br}
\affiliation{Departamento de F\'{\i}sica, ICEX, Universidade Federal de 
Minas Gerais, CP 702, 30123-970,Belo Horizonte, MG, Brazil.}

\date{\today}

\begin{abstract} 
{\it Ab initio} calculations indicate that topological-defect networks
in graphene display the full variety of single-particle electronic
structures, including Dirac-fermion null-gap semiconductors, as well
as metallic and semiconducting systems of very low formation energies
with respect to a pristine graphene sheet. Corrugation induced by the
topological defects further reduces the energy and tends to reduce the
density of states at the Fermi level, to widen the gaps, or even to lead to
gap opening in some cases where the parent planar geometry is metallic.

\end{abstract}


\maketitle

Since its recent experimental discovery~\cite{novos04}, graphene has emerged
as an unique condensed-matter system, for its remarkable electronic
and electromechanical properties, as well as a test-bed for
relativistic-dynamics phenomena, due to the linear and isotropic
dispersion of the electronic states near the Fermi level, that leads
to the massless-Dirac-fermion nature of its low-energy electronic
excitations~\cite{rmp}. Devising gap-opening and gap-control
mechanisms is one of the key issues related to the prospective
application of graphene as the new holy-grail of electronic
devices~\cite{rmp,oostinga}. Several mechanisms for changing, perhaps
tailoring, the electronic structure of graphene have been
proposed. Experimental and theoretical works have indicated the
occurrence of gap opening at the Dirac point in graphene due to
interaction with insulating substrates~\cite{szhou}. The generation of
additional Dirac cones of massless fermions has also been identified,
due either to imposed periodic perturbations~\cite{chpark}, or
associated with a periodic pentagon-heptagon superstructure on grain
boundaries in polycrystalline graphene~\cite{joice}. Of no smaller
relevance are the effects of corrugation on the thermodynamic
stability and transport properties of substrate-supported and
free-standing graphene~\cite{rmp,lui,meyer,fasolino,varchon}. What
causes corrugation in graphene is an open issue: the possibility of
corrugation being an intrinsic feature of the
material~\cite{meyer,fasolino}, due to the interaction with a
substrate~\cite{varchon}, or the effect of adsorbates~\cite{rebecca},
are scenarios that have been put forward in the literature.

In this work, we employ {\it ab initio} calculations to examine the
possibility of topological curvature, its stability and energetics,
and its effects on the electronic structure of graphene
sheets. Topological curvature fields are generated when, following the
formation of a Stone-Wales defect (SW) by a 90$^\circ$ rotation of a
carbon-carbon bond~\cite{crespi}, the two pentagons and two heptagons
comprising the SW are separated, and the defect is fully
dissociated. We consider periodic lattices containing two pentagons
and two heptagons, such that these isolated odd-membered rings develop
local curvature fields, while preserving the null average curvature of
the parent graphene sheet. Our study broadens the scope of the
pentaheptite allotrope of carbon previously introduced by Crespi and
collaborators~\cite{crespi}, that consisted of a two-dimensional (2D)
carbon sheet composed entirely of pentagons and heptagons. Our results
raise the possibility of the generation of semiconducting and metallic
graphene sheets by the introduction of topological defects (TD). We
find that periodic networks of TDs may present the full variety of
electronic structures, at the single-particle level, including genuine
metals, ``graphene-like'' null-gap semiconductors, and also finite-gap
($\sim$0.02-0.75~eV) semiconductors. A remarkable result of our
calculations is the identification of metallic and semiconducting
structures with very small formation energies ($\sim$0.1~eV/atom with
respect to graphene), in which the corrugation induced at the TD sites
plays a key role in reducing the energy. Being a full first principles
address of {\it fully dissociated} TDs, our work differs from previous
theoretical works on corrugated graphene and on topological defects in
this material~\cite{paco,vozmed}. In particular, our results cast doubt on
the assumption that Dirac points are robust against the introduction
of a distribution of TDs in graphene~\cite{vozmed}.

Our calculations are performed in the framework of Kohn-Sham density
functional theory (DFT), within the generalized-gradient approximation
(GGA)~\cite{ks-gga} and norm-conserving pseudopotentials in the
Kleinman-Bylander factorized form~\cite{tm-kb}. We use the LCAO method
implemented in the SIESTA code~\cite{siesta}, with a double-zeta basis
set plus polarization orbitals. Calculational parameters are set such
that total-energy differences are converged to better than 1~meV/atom,
with residual atomic forces no larger than
0.02~eV/\AA. Single-particle energy eigenvalues are converged to
$\sim$1~meV.

The pentaheptite form of graphene investigated by Crespi {\it et
al.}~\cite{crespi} is shown in Fig.~\ref{fig1}(a). The structures we
investigate in this work, which can be considered a generalization of
the pentaheptite, are generated from three different seed structures,
as shown in Fig.~\ref{fig1}. The first seed in Fig.~\ref{fig1}(a) is
the pentaheptite itself, with eight atoms in the unit cell. The second
seed, in Fig.~\ref{fig1}(b), has a 16-atom unit cell with a SW
surrounded by hexagons. The third seed, shown in Fig.~\ref{fig1}(c),
has a 24-atom unit cell where the pentagons and heptagons are already
dissociated, each being fully surrounded by one shell of
hexagons. Larger unit cells can be constructed from the seeds, by
introducing shells of hexagons surrounding the TDs. In order to
establish a nomenclature, let the 8-atom seed be called $S_{11}$, with
a TD number density $n(1,1)= 0.18$ (in units of \AA$^{-2}$). The
16-atom seed is then labeled $S_{21}$, and the 24-atom seed is
$S_{31}$. From these we can generally build an $S_{ij}$ geometry which
is the $j$-th member of the $i$-th family, having an $8 \times i\times
j^2$-atom unit cell and a TD number density of $n(i,j) = n(1,1)/(i
\times j^2)$. Here, we consider a total of nine planar geometries, of
which: four are generated from the 8-atom seed, namely $S_{11}$,
$S_{12}$, $S_{13}$, and $S_{14}$; three are built from the 16-atom
seed, namely $S_{21}$, $S_{22}$, and $S_{23}$; and two from the
24-atom seed, namely $S_{31}$ and $S_{32}$~\cite{s33}.

For all but the $S_{11}$ and $S_{21}$ geometries, the SW
defects are fully dissociated, and corrugated structures can be
generated by allowing for the local curvature fields generated by the
isolated TDs to develop. Thus, a total of 16 structures were
considered, nine of which are planar and seven are
corrugated. Figure~\ref{fig2} shows a 2x2 unit cell of the planar and
the corrugated $S_{23}$ geometries. In all cases, the planar
geometries did not develop corrugation spontaneously under structural
relaxation. In order to investigate whether this was an indication of
metastability of the planar forms, we built slightly corrugated
initial geometries with random off-plane shifts of the atoms imposed
on the planar structures. Upon structural optimization the
$S_{31}\;\left[n(3,1) = 0.060\right]$, $S_{12}\;\left[n(1,2) =
0.045\right]$, and $S_{22}\;\left[n(2,2) = 0.023\right]$ geometries
relaxed back to the planar form for off-plane shifts of 0.01~\AA, and
developed corrugation for larger shifts, with the exception of the
$S_{31}$, with a higher TD concentration, that only developed
corrugation for shifts of 0.1~\AA\ or larger. Thus, we find numerical
evidence for a metastability of the planar structures with high TD
concentrations, down to a critical density of TDs ($n_c \sim 0.023$),
below which a corrugation instability sets in, as indicated in
Fig.~\ref{fig3}. The instability is related to the fact that, for
the more diluted structures, large incipient conical and saddle-like
regions are associated with the isolated pentagons and heptagons,
respectively, and the cost of flattening such regions onto the plane
becomes correspondingly larger~\cite{stable}.

In Fig.~\ref{fig3} we show, for the 16 structures, the formation
energy (with respect to pristine graphene) $E_f = E_{tot}/N -
E_{graph}$ (in eV/atom) as a function of the TD number density, where
$N$ is the number of atoms in the unit cell, $E_{tot}$ is the the
total energy per unit cell, and $E_{graph}$ is the total energy per
atom for a pristine graphene sheet. Since the planar sheets remained
numerically stable, we include in the figure the energy of their
``relaxed'' geometries. Clearly, dissociation of TDs, when constrained
to the plane, incurs an energy cost that behaves linearly with $n$
(with a negative slope), with the exception of the $S_{14}$, a
structure in the limit of the connectivity of such geometries in the
plane~\cite{s33}. Corrugation leads to a reduction of the formation
energy that is comparable to or larger than the formation energy
itself for all but the $S_{31}$ and $S_{12}$ geometries. We define a
corrugation energy by the difference between the formation energies of
the planar and corrugated geometries $E_{c} = E_{f}^{plane} -
E_{f}^{corr}$. $E_f$ and $E_{c}$ values are included in Table I. Note
the correlation between $E_{f}$ of the planar geometries and $E_{c}$
in Fig.~\ref{fig3}: the more diluted structures have large $E_f$
values in the planar form and also large values of $E_{c}$, leading to
a trend, for the corrugated geometries in each family, of decreasing
formation energies as the TD concentration decreases.

Of note also is the fact that the $S_{2j}$ family is a low energy
branch in Fig.~\ref{fig3}, both for planar and corrugated forms,
while the $S_{1j}$ and $S_{3j}$ families have comparable
energies. Among the seed geometries, the $S_{21}$ has a much lower
formation energy than the $S_{11}$ and $S_{31}$. This can be
understood by comparing the seed structures: the $S_{11}$ seed has
zero hexagons and four TDs in the unit cell; the $S_{21}$ has four
hexagons and four TDs, and the the $S_{31}$ has eight hexagons and
four TDs. In the latter, the hexagons are highly strained due to the
SW dissociation, and this structure has the highest $E_{f}$ among the
seeds. The $S_{21}$ has four slightly strained hexagons in its
periodic unit, and this considerably lowers its energy when compared
with the $S_{11}$ and $S_{31}$. As a result of the balance between
hexagons and TDs in the seed, the bond-stretching and bond-bending
distortions in the $S_{2j}$ family are smaller than in the other two
families. To quantify these structural features, we compute the
Keating-model elastic energies of the 16 structures. The results are
included in Fig.~\ref{fig3} and in Table~I.  Note that the {\it
ab initio} $E_f$ and the Keating energies follow the same overall
trends, with the $S_{2j}$ family appearing as a low-elastic-energy
branch in Fig.~\ref{fig3}.

Also related to these structural features of the seeds is the degree
of corrugation of each structure. We define the atomic corrugation
$Z_c$ of the non-planar structures as the distance of the center of
the protruding pentagons to the average plane. The values of $Z_c$ are
given in Table~I, and the inset in Fig.~\ref{fig3} shows the
behavior of $Z_c$ as a function of $n$. Note that the $S_{2j}$ family
represents a low corrugation branch, due to its comparably less
strained planar geometry than the corresponding ones in the $S_{1j}$
and $S_{3j}$ families. This also explains why the planar $S_{22}$ with
$n(2,2) = 0.023$ is metastable, while the the planar $S_{13}$ with
$n(1,3) = 0.020$ is unstable against corrugation.

The diverse nature of the electronic states of these TD periodic
lattices can be anticipated by examining the DOS plots of the seed
structures in Fig.~\ref{fig1}. The DOS for the $S_{11}$ seed, in
Fig.~\ref{fig1}(d), displays a metallic character, with an {\it ab
initio} value for the DOS at the Fermi level that is half of the
tight-binding result in Ref.~\onlinecite{crespi}. The $S_{21}$ seed,
in Fig.~\ref{fig1}(e), shows a ``graphene-like'' Dirac-fermion DOS
that derives from linear electronic bands crossing each other at the
Fermi point, forming a Dirac cone on a much reduced energy scale
($\sim$[-0.1,0.1]~eV), beyond which other bands intervene and break
the electron-hole symmetry. The $S_{31}$ seed is a semiconductor [DOS
is shown in (f)] with a GGA-DFT gap of $\sim$0.6~eV.

\begin{table}[h]
\vspace{-0.4cm}
\caption{Physical properties of TD networks in graphene. Formation
energy $E_f$, corrugation energy $E_c$ (see text), and elastic
energies $E_{el.}$ are in units of eV/\AA.  $Z_c$ (in \AA) is the
measure of corrugation (see text). $E_g$ (in eV) is the energy gap for
the semiconducting systems. The value of DOS at the Fermi level (in
number of states per energy per atom) is given for the metallic
systems. The accumulated charge on the pentagons $Q_{P}$ is given for
the corrugated geometries.}
\begin{tabular} {|c|c|c|c|c|c|c|c|c|c|c|}
\hline
  &\multicolumn{4}{|c|}{planar} \vline &\multicolumn{6}{|c|}{corrugated} \\ [0.5ex]
  \cline{1-11}
 $S_{ij}$ &$E_f$ &$E_{el.}$ &$E_g$ &DOS &$E_f$ &$E_{c}$ &$Z_{c}$ &$E_g$ &DOS &$Q_P$\\
\hline
$S_{11}$ &0.24 &0.32 &  -  &0.05 &  -  &  -  &  -  &  -  &  -  & - \\
\hline
$S_{12}$ &0.40 &0.38 &  -  &0.07 &0.35 &0.05 &1.03 &  -  &0.04 &-0.016 \\
\hline
$S_{13}$ &0.44 &0.42 &0.53 & -   &0.21 &0.23 &2.58 &0.67 &  -  &-0.015 \\
\hline
$S_{14}$ &0.46 &0.44 &0.02 & -   &0.15 &0.31 &4.03 & -   &0.02 &-0.015 \\
\hline
$S_{21}$ &0.14 &0.18 &0.00 &0.00 &  -  &  -  &  -  &  -  &  -  & - \\
\hline
$S_{22}$ &0.26 &0.24 &  -  &0.03 &0.16 &0.10 &1.80 &0.03 &  -  &-0.009 \\
\hline
$S_{23}$ &0.29 &0.26 &0.42 &  -  &0.09 &0.20 &3.29 &0.39 &  -  &-0.012\\
\hline
$S_{31}$ &0.36 &0.37 &0.61 &  -  &0.33 &0.03 &0.58 &0.74 &  -  &0.000\\
\hline
$S_{32}$ &0.47 &0.44 &0.00 &0.00 &0.17 &0.30 &3.28 &0.50 &  -  &-0.013\\
\hline
\end{tabular}
\vspace{-0.2cm}
\end {table}

Regarding the general aspects of the electronic structure of all
planar and corrugated structures in our study, a first observation
concerns the role played by topology, relaxation, and corrugation, in
determining the nature of the electronic states. In order to address
the effects solely of the topological transformation (TT), we compute
the DOS and the electronic bands of the (planar) seed structures in
three different situations, as also shown in Fig.~\ref{fig1}: (1) a
structure with only the TT imposed on graphene, without any relaxation
of either atomic positions or lattice vectors; (2) with the TT and
relaxation of atomic positions only; (3) and finally, with the TT and
also relaxation of the lattice vectors and atomic positions. For the
$S_{11}$, in Fig.~\ref{fig1}(d), relaxation does not bring any
significant changes, and the system is metallic in the three cases,
with little change in the position of the Fermi level. Things are
quite different for the $S_{21}$, in Fig.~\ref{fig1}(e), which is a
metallic system in the first two scenarios above, but develops a
Dirac-fermion-like band structure when lattice relaxation is allowed,
with the Fermi level lying on the Dirac point. For the $S_{31}$ we
only computed the electronic states for the cases (1) and (3). The
DOS's are shown in Fig.~\ref{fig1}(f), where case (1) is seen to
behave as a metal, with the Fermi level lying at the bottom part of a
set of bands that start above an energy gap of about 0.5 eV. After
full relaxation, enough spectral weight is shifted to lower energies
for the Fermi level to lye on the gap, and the system to become a
semiconductor. Hence, the TT by itself does not explain the behavior
of the electronic states we observe in our calculations, with
relaxation effects playing a crucial role. The realistic treatment of
the interatomic interactions, provided by the {\it ab initio} method,
is thus a key ingredient.

In Table~I we include the values of the gap for all semiconducting TD
networks, and the value of the DOS at the Fermi level for metallic
ones. Generally, corrugation tends to reduce the DOS at the Fermi
level, to widen the gap, or even lead to gap opening in some cases
where the parent planar geometry is metallic. There is one exception,
however, in the case of the $S_{14}$ which displays a small gap
(0.02~eV) in the planar form and is a metal in the corrugated
geometry. Overall, a total of nine structures are semiconductors,
seven of which with gaps ranging from 0.39 to 0.74~eV.  Five of the
metastable systems in our study have formation energies below
0.2~eV/atom ($E_f = 0.39$~eV/atom for a fullerene molecule with the
same methodology). Among these, the planar $S_{21}$ is a Dirac-fermion
null-gap semiconductor, the corrugated $S_{14}$ is a metal, and the
corrugated $S_{22}$, $S_{23}$, and $S_{32}$ are semiconductors. The
corrugated $S_{23}$ TD network stands out as a semiconductor with very
small formation energy (0.09~eV/atom) for a carbon system.

The low formation energies of the corrugated TD networks in our study,
as well as the $Z_c$ values which are in the range of the corrugation
observed experimentally~\cite{rmp,lui,meyer,fasolino,varchon}, suggest
that consideration should be given to these structures as a model for
the rippled graphene samples in the experiments. A feature of
corrugated graphene that has been experimentally observed is the
formation of electron-hole puddles, with a concentration of electrons
in the higher parts of the samples~\cite{parga}. In our corrugated
geometries we find that the pentagons at the higher parts of the
ripples are electron-rich (except for the corrugated $S_{31}$) as
shown in Table~I. Moreover, the low formation energies indicate that
the synthesis of such structures may be feasible. We anticipate that
kinetics would be the only potentially limiting factor, since besides
the low formation energies, the barriers protecting such structures
from reverting back to pristine graphene should be of the order of the
7~eV estimate for the $S_{11}$ in Ref.~\onlinecite{crespi}. Regarding
a pressure-driven transformation of graphite into a three-dimensional
bulk of TD networks, as envisaged in Ref.~\onlinecite{crespi}, we note
that the areal density of all corrugated structures in our study is
larger than that of graphene, e.g., one carbon atom per 2.35~\AA$^2$
for the $S_{23}$ compared with one carbon atom per 2.69~\AA$^2$ for
graphene, which would favor a pressure-driven transformation. We add
that very recent experiments have indicated the formation of both
clustered and isolated pentagon-heptagon pairs in graphene samples
obtained by reduction of graphene-oxide~\cite{cristina}, and
corrugation was observed in the areas of the samples containing the
defects. Further refinement of such reduction techniques may lead to a
viable synthesis routes for TD networks. Moreover, this also suggests
that other chemistry-based synthesis techniques may also be
possible~\cite{crespi}.

In summary, {\it ab initio} calculations indicate that the electronic
structure of topological-defect networks in graphene range from
Dirac-fermion null-gap semiconductors to finite-gap semiconductors and
metals, some of them with very low formation energies, offering a
possible route for tailoring the electronic nature of graphene
sheets. Corrugation induced by the topological defects further reduces
the energy and modifies the electronic structure of the sheets.

\begin{acknowledgments}
We thank Jos\'e M. Soler for useful discussions, and acknowledge
support from CNPq, FAPEMIG, and Instituto do Mil\^enio em
Nanoci\^encias-MCT, Brazil. 
\end{acknowledgments}

\begin{figure}[t]
\caption{(Color online) (a-c): 2x2 unit cells of seed structures. (a)
$S_{11}$, (b) $S_{21}$, and (c) $S_{31}$. Primitive-cell vectors are
indicated.  (d-f): DOS for the structures (a) to (c). The DOS's are
depicted for the structures with only the topological transformation
(dashed red lines), with relaxation of atomic positions (dot-dashed
blue lines), and with relaxation of atomic positions and lattice
vectors (full black lines).}
\vspace{-0.6cm}
\label{fig1}
\end{figure}

\begin{figure}[t]
\caption{(Color online) (a) Planar and (b) corrugated forms of the 
$S_{23}$, the lowest-energy topological-defect network.}
\vspace{-0.6cm}
\label{fig2}
\end{figure}

\begin{figure}[t]
\caption{(Color online) {\it Ab initio} formation energy of planar
(filled) and corrugated (empty symbols) TD networks, as a
function of TD density. Symbols with a diagonal stripe
filling show the Keating-model elastic energies of the planar
geometries. Planar structures to the right of the dashed vertical
line are metastable. Inset shows corrugation height as a function of
TD density. Lines are guides to the eye.}
\vspace{-0.6cm}
\label{fig3}
\end{figure}


\begin{thebibliography}{100}
\bibitem{s33} For structures with smaller TD densities and thus larger
incipient conical and saddle-like regions, it is not possible to build
a fully threefold-connected network, with reasonable bond-length
values, without introducing corrugation. In practice, we found this to
be true already for a planar $S_{33}$ structure, with $n(3,3) =
0.007$, for which we were not able to seam the tears that open when
additional shells of hexagons are introduced between the TDs.
\bibitem{stable} In Ref.~\onlinecite{crespi} it was suggested that
sufficiently homogeneous distributions of equal numbers of pentagons
and heptagons should be stable in the planar form. Our calculations
indicate that, generally, too dilute distributions ($n<n_c$) should be
unstable against corrugation.
\end{thebibliography}
\end{document}